\documentclass[floatfix,reprint,prb,superscriptaddress,twocolumn]{revtex4}
\usepackage{graphicx}
\usepackage{amsmath}
\usepackage{amsfonts}
\usepackage{amssymb}
\usepackage{amsbsy}
\usepackage{hyperref}

\begin{document}
\title{Magnetic fan structures in Ba$_{0.5}$Sr$_{1.5}$Zn$_2$Fe$_{12}$O$_{22}$ hexaferrite revealed by resonant soft X-ray diffraction}
\date{\today}

\author{Alexander J. Hearmon}
\email{a.hearmon@physics.ox.ac.uk}
\affiliation{Clarendon Laboratory, Department of Physics, University of Oxford, Parks Road, Oxford OX1 3PU, United Kingdom}
\affiliation{Diamond Light Source Ltd., Harwell Science and Innovation Campus, Didcot, OX11 0DE, United Kingdom}

\author{R. D. Johnson}
\affiliation{Clarendon Laboratory, Department of Physics, University of Oxford, Parks Road, Oxford OX1 3PU, United Kingdom}
\affiliation{ISIS Facility, STFC-Rutherford Appleton Laboratory, Didcot, OX11 OQX, United Kingdom}

\author{T. A. W. Beale}
\affiliation{Department of Physics, Durham University, South Road, Durham, DH1 3LE, United Kingdom}
 
\author{S. S. Dhesi}
\affiliation{Diamond Light Source Ltd., Harwell Science and Innovation Campus, Didcot, OX11 0DE, United Kingdom}

\author{X. Luo}
\affiliation{Laboratory for Pohang Emergent Materials and Department of Physics, Pohang University of Science and Technology, Pohang 790-784, Korea}

\author{S.-W. Cheong}
\affiliation{Laboratory for Pohang Emergent Materials and Department of Physics, Pohang University of Science and Technology, Pohang 790-784, Korea}
\affiliation{Rutgers Center for Emergent Materials and Department of Physics and Astronomy, Piscataway, New Jersey 08854, USA}

\author{P. Steadman}
\affiliation{Diamond Light Source Ltd., Harwell Science and Innovation Campus, Didcot, OX11 0DE, United Kingdom}

\author{Paolo G. Radaelli}
\affiliation{Clarendon Laboratory, Department of Physics, University of Oxford, Parks Road, Oxford OX1 3PU, United Kingdom}

\begin{abstract}
The hexaferrites are known to exhibit a wide range of magnetic structures, some of which are connected to important technological applications and display magnetoelectric properties. We present data on the low magnetic field structures stabilised in a Y-type hexaferrite as observed by resonant soft X-ray diffraction. The helical spin block arrangement that is present in zero applied magnetic field becomes fan-like as a field is applied in plane. The propagation vectors associated with each fan structure are studied as a function of magnetic field, and a new magnetic phase is reported. Mean field calculations indicate this phase should stabilise close to the boundary of the previously reported phases. 
\end{abstract}
\maketitle

\section{Introduction}
\noindent
Hexaferrites have been studied with great interest for many years as the different structural types and the possibility of tuning their behaviour with doping (varying the relative strengths of the exchange interactions) has lead to a large degree of flexibility in the number of different magnetic phases supported by this class of materials \cite{gorter1957saturation_265,Enz1961magnetization_170}. Their room temperature magnetism has also given rise to use in numerous applications, and their layered structure makes them suitable for growth by epitaxial methods \cite{glass1977growth_416}. Further exotic magnetic arrangements can typically be induced in these materials by altering the temperature and applied magnetic field. In particular, there has been substantial interest recently in the Z-type hexaferrites [formula (Ba,Sr)$_3M_2$Fe$_{24}$O$_{41}$ where $M$ is a divalent metal ion], which show the magnetoelectric effect at room temperature \cite{kitagawa2010lowfiled_50}, as well as the Y-type hexaferrites. Of the latter, two compositions in particular have been well-studied: (i) Ba$_2$Mg$_2$Fe$_{12}$O$_{22}$ and (ii) Ba$_{2-x}$Sr$_x$Zn$_2$Fe$_{12}$O$_{22}$. The Mg system displays helical, conical, and ferrimagnetic arrangements \cite{Momozawa1986magnetic_99,ishiwata2010neutron_142} as the applied magnetic field is varied, and a ferroelectric polarisation appears in the magnetic field induced `tilted conical' phase that can be controlled with the direction of the applied field \cite{ishiwata2008helimagnet_184,taniguchi2008ferroelectric_168,sagayama2009two_169}. It was also shown recently that this system displays signatures of an electric-dipole-active magnetic resonance (`electromagnon') measured by optical techniques \cite{Kida2009electric_173}. The magnetic structure of Ba$_{2-x}$Sr$_x$Zn$_2$Fe$_{12}$O$_{22}$, in contrast to the Mg system, evolves from a helimagnetic arrangement in zero applied magnetic field to various `fan' structures as the transverse field is increased \cite{momozawa1985magnetic_174A,momozawa1993field_26}. The discovery by Kimura \emph{et al.~}in 2005 that one of these fan structures also exhibits ferroelectricity \cite{kimura2005electric_24} (this has also been reported recently in aluminium doped systems where a conical component of the magnetic structure develops \cite{lee2011field_195,Chun2010giant_206}) has sparked renewed interest in practical applications since these systems have been shown to display multiferroic properties at room temperature. Electron diffraction experiments have also shown that the various magnetic structures induce modulations in the crystal lattice \cite{asaka2011lattice_190}. However, the origin of the magnetic field induced ferroelectric polarization in Ba$_{2-x}$Sr$_{x}$Zn$_2$Fe$_{12}$O$_{22}$ remains a mystery, since all magnetic structures so far determined by neutron diffraction are centrosymmetric. 

Resonant soft X-ray diffraction is an excellent tool with which to study the magnetism in these hexaferrites because, by tuning the X-ray energy to be resonant with the Fe L-edge, a substantial enhancement in the magnetic scattering signal can be obtained. This was recently employed by Mulders \emph{et al.~}\cite{mulders2010circular_141} where circularly polarised X-rays were used to examine the zero field helical structure. The technique was subsequently used to great effect by Hiraoka \textit{et al.~}\cite{Hiraoka2011spin_248} where the contrast in the diffracted signals from the helical spin arrangement between right and left circularly polarised X-rays was exploited to map the spatial distribution of chiral domains present in these materials. What makes resonant soft X-ray diffraction particularly suited to this problem is its sensitivity to different projections of the magnetic structure factor through different elements of the polarisation matrix, making it capable, at least in principle, of detecting subtle deviations from quasi-centrosymmetric magnetic structures. The technique also has high reciprocal space resolution and consequently is very sensitive to changes in the magnetic propagation vector(s) associated with the different phases of the material. In this paper we present a detailed diffraction study of the magnetic structures present in Ba$_{0.5}$Sr$_{1.5}$Zn$_2$Fe$_{12}$O$_{22}$, in the low applied magnetic field region of the phase diagram.  We discuss the various phases that are observed in the data and construct an outline magnetic field / temperature phase diagram for the system. In addition we report a new `6-fan' magnetic structure that gives rise to magnetic scattering with propagation vector $\mathbf{q} = (0,0,1)$, and energy calculations are carried out to determine a candidate spin arrangement for this new phase. Thus, in making the transition from helimagnet in zero applied field to ferrimagnet in high fields, the hexaferrite explores a variety of intermediate fan states with different periodicities, but similar net magnetisations and energies in the mean field model. 

The Y-type hexaferrites have a relatively complex crystal structure \cite{Braun1957_165} with a large unit cell: the space group is $R\bar{3}m$ and the lattice parameters of the sample used in the synchrotron experiment were measured at room temperature using a laboratory `SuperNova' X-ray source to be $a = 5.852(6)$~{\AA} and $c = 43.54(4)$~{\AA} (the correspondingly small size of $c^*$ makes the long X-ray wavelengths available with soft X-ray experiments particularly appropriate for the study of this system). The structure consists of an alternate stacking of spinel `short' blocks (conventionally referred to as `S' blocks), and hexagonal `long' blocks (`L' blocks). The doping $x$ measures the relative amounts of Ba and Sr, and a further parameter $\gamma$ is necessary to describe the mixing of Fe and Zn on the $6c$ sites: it is given by the fraction of Fe on the $6c$ sites in the L block (which is equal to the fraction of Zn on the 6c sites in the S block) \cite{Novak2007magnetism_171}. In ref.~\onlinecite{momozawa1985magnetic_174A} a value of $\gamma = 0.661$ is reported for $x$ close to 1.5, and this is the value that is used in describing the moments of the spin blocks here.   

\begin{figure}[tb]
\begin{center}
\includegraphics[scale=0.4]{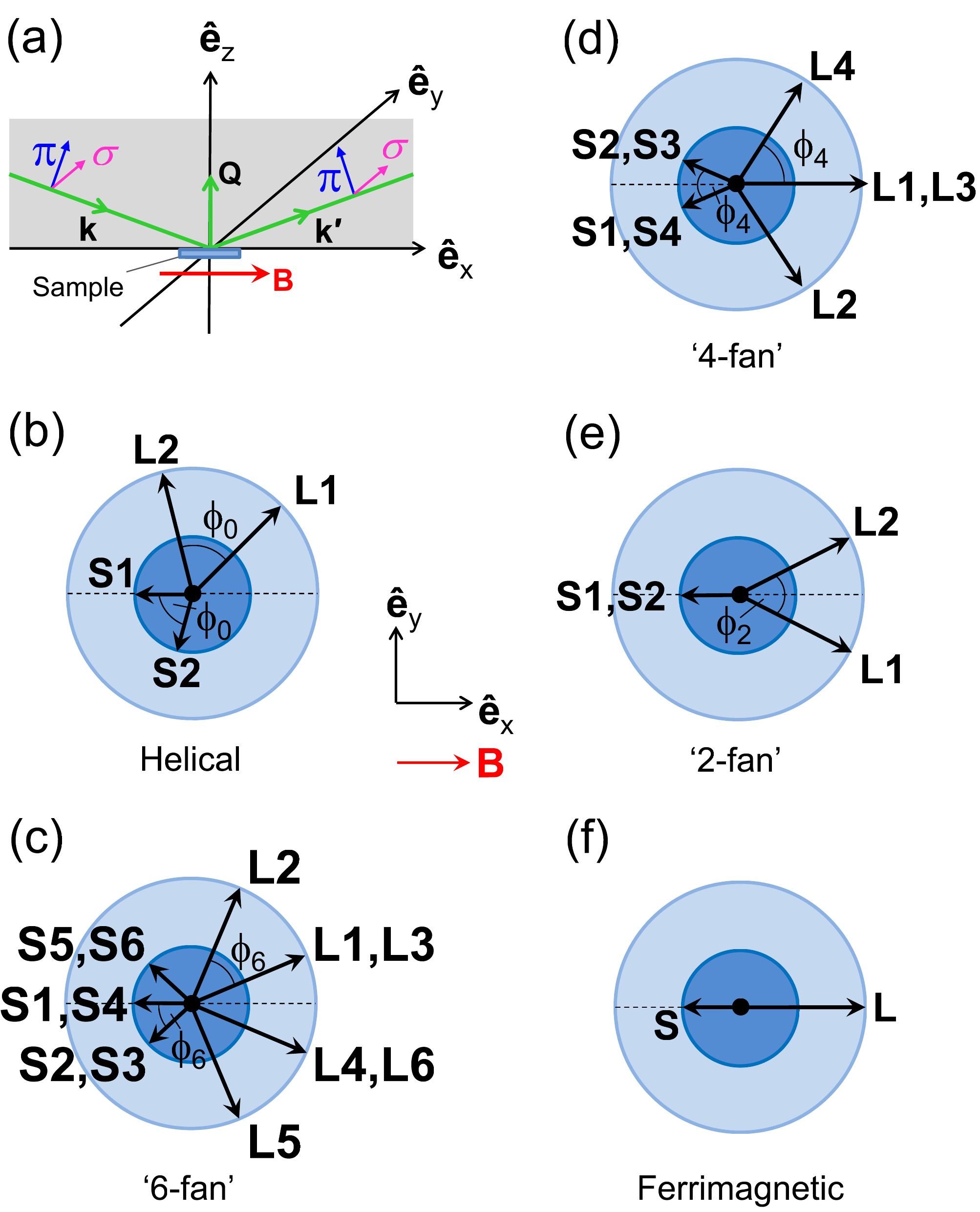}
\caption{(Color online) (a) Scattering geometry for the resonant soft X-ray diffraction experiment. The scattering vector $\mathbf{Q} = \mathbf{k}^\prime - \mathbf{k}$ is directed along $\mathbf{c}^*$, and the applied magnetic field is orthogonal to this and within the scattering plane. The directions of $\sigma$ and $\pi$ linear X-ray polarisations are indicated. (b-f) Plan views of the various configurations of the magnetic L and S blocks as a function of increasing applied field \cite{momozawa1993field_26}. Angles between adjacent L (or S) blocks are indicated. The blocks stack in the order S1-L1-S2-L2-$\dots$ as one moves along the positive $\mathbf{c}$ direction. The Cartesian basis vectors $\hat{\mathbf{e}}_{x,y,z}$ and the applied field direction, $\mathbf{B}$, are indicated.   
\label{Hexaferrite_Paper_Structures}}
\end{center}
\end{figure}

\begin{table}[tb]
\begin{center}
\caption{The propagation vectors associated with the various helical and fan magnetic structures of fig. \ref{Hexaferrite_Paper_Structures}. The final column indicates those magnetic satellites observable at the resonant energy in the $\sigma \rightarrow \pi^\prime$ channel with the experimental geometry depicted in fig. \ref{Hexaferrite_Paper_Structures}(a). Bragg peaks are subject to the selection rule $-h + k + l = 3n$ in the rhombohedral setting. \label{HEXA_tab:Satellites}}
\vspace{0.15cm}
\begin{tabular}{lclcl}
\hline \hline
Phase	&\hspace{0.25cm}  	&  Propagation 													&\hspace{0.25cm} 	& Magnetic reflections    \\
		&  					&  vectors, $\mathbf{q}$										&  					& observed in $ \sigma \rightarrow \pi^\prime $ \\
\hline \hline
Helical &  &  $(0,0,q_\textrm{helix})$									&  &	$(0,0,3 \pm q_\textrm{helix})$	 	 \vspace{0.15cm}	 \\
\hline
6-fan	&  &  $(0,0,0.5)$ 												&  & 	$-$	\\
		&  &  $(0,0,1)$	 												&  &   $(0,0,2)$ and $(0,0,4)$			\\
		&  &  $(0,0,1.5)$	 											&  &   $-$		\\
		&  &  $(0,0,0)$			 										&  &   $(0,0,3)$			\vspace{0.15cm}   \\
\hline
4-fan	&  & $(0,0,0.75)$												&  & 	 $-$	\\
		&  & $(0,0,1.5)$												&  & 	 $(0,0,1.5)$ and $(0,0,4.5)$	\\
		&  & $(0,0,0)$													&  & 	 $(0,0,3)$ 	\vspace{0.15cm}	\\
\hline
2-fan	&  &  $(0,0,1.5)$												&  & 	 $-$			 \\
 		&  & $(0,0,0)$													&  & 	 $(0,0,3)$	\\
\hline \hline
\end{tabular}\end{center}
\end{table}

\begin{figure*}[t]
\begin{center}
\includegraphics[scale=0.55]{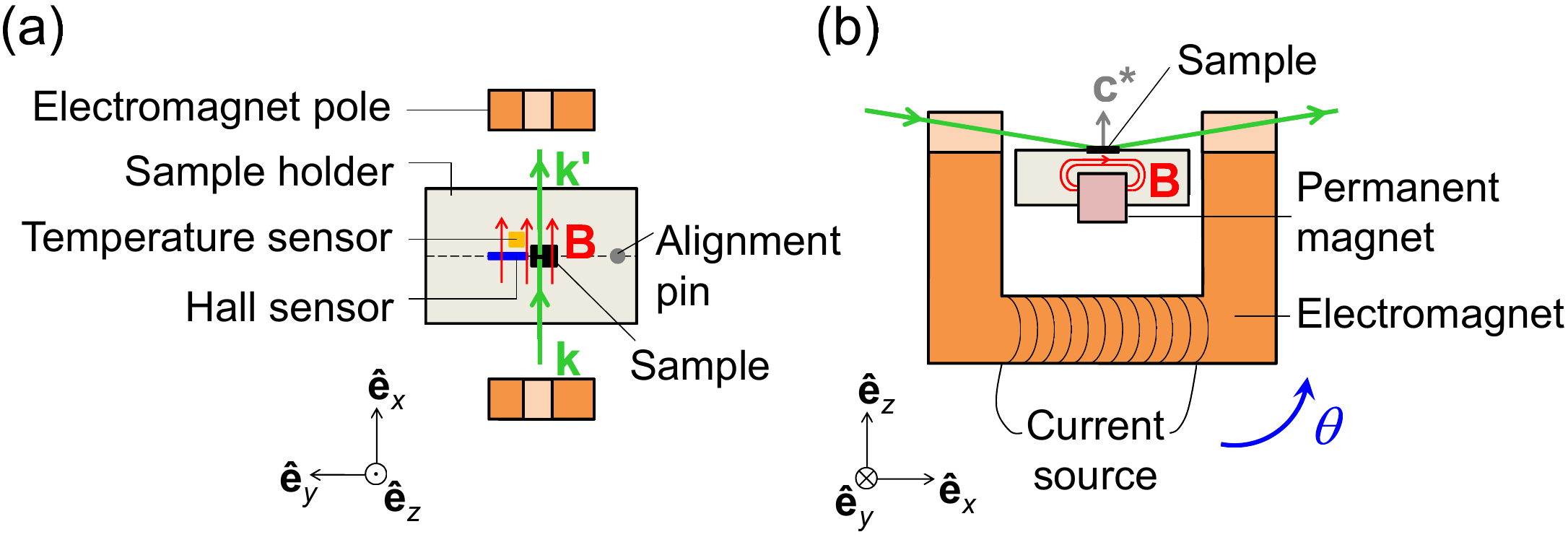}
\caption{(Color online) The RASOR sample mount seen in two different orientations (the $\hat{\mathbf{e}}_{x,y,z}$ directions, as defined in figure \ref{Hexaferrite_Paper_Structures}, are indicated). The electromagnet has two central sections removed to allow the X-ray beam to access the sample, and the entire assembly is rigidly attached to the $\theta$ arm and rotates with it, the positive sense being depicted by the arrow in (b).\label{Hexaferrite_Paper_SampleMount}}
\end{center}
\end{figure*}

Various magnetic structures, determined by neutron scattering in ref.~\onlinecite{momozawa1993field_26}, have been reported to date. These systems have many different sites containing magnetic ions, and as such the possibility exists for a large number of complex and varied magnetic structures to appear in the material. However, a substantial simplification can be made because all of the magnetic moments within each L or S block are ferromagnetically aligned perpendicularly to $\mathbf{c}$, such that the spin blocks can be treated essentially as separate effective magnetic moments stacked along $\mathbf{c}$. The magnetic phases that the material exhibits as a function of increasing applied magnetic field are summarised in fig. \ref{Hexaferrite_Paper_Structures}. In zero field an incommensurate helical structure is stabilised [fig. \ref{Hexaferrite_Paper_Structures}(b)], with the S blocks and L blocks out of phase by an angle of $180^\circ + \phi_0/2$. Applying a transverse field causes the structure to change such that more of the moment is aligned with the field: this results in the `Intermediate-I' or `4-fan' oscillatory structure of fig.~\ref{Hexaferrite_Paper_Structures}(d), and at higher fields the `Intermediate-II' (`2-fan') structure of fig.~\ref{Hexaferrite_Paper_Structures}(e). Note that a strong antiferromagnetic exchange interaction between nearest neighbour spin blocks tends to align the S blocks antiparallel to the field. As the field is further increased, the system enters the multiferroic `Intermediate-III' phase in which a ferroelectric polarisation is reported \cite{kimura2005electric_24}, and at still higher fields the system becomes ferrimagnetic [fig.~\ref{Hexaferrite_Paper_Structures}(f)]. Note that evidence for a `6-fan' phase of fig.~\ref{Hexaferrite_Paper_Structures}(c) has not been previously reported but is necessary to interpret our data, as will be discussed later. The propagation vectors describing the different phases can be calculated via the standard Fourier expansion of the collection of moments $\boldsymbol{\mu}_l$ from which the structure is generated:
\begin{equation}
\boldsymbol{\mu}_l \propto \sum_j \left( \mathbf{S}_j e^{-i\mathbf{q}_j\cdot \mathbf{r}_l} + \mathbf{S}_j^* e^{i\mathbf{q}_j\cdot \mathbf{r}_l}  \right),
\end{equation}
where $\mathbf{S}_j$ is the Fourier component for the $j$th propagation vector, and $\mathbf{r}_l$ is the position of the moment $\boldsymbol{\mu}_l$. The propagation vectors associated with each structure are given in table \ref{HEXA_tab:Satellites}. It is known (see, for example, ref.~\onlinecite{kimura2005electric_24}) that the system very quickly changes from one phase to another as the applied field is increased. Thus, the motivation of the present work is to obtain a detailed field dependence of the system in the low applied field part of the phase diagram, where the energies of the different phases are closely spaced.

\section{Experimental details}
\noindent
Single crystals of Ba$_{0.5}$Sr$_{1.5}$Zn$_2$Fe$_{12}$O$_{22}$ were grown using a flux technique with Na$_2$O-Fe$_2$O$_3$. After being melted at $1420~^\circ$C, the flux mixture was subjected to several thermal cycles to obtain the right phase, and then slowly cooled to room temperature \cite{Momozawa1987403}. The single crystals are hexagonal plates and were polished with the $c$-axis normal. The resonant soft X-ray diffraction experiment was carried out at beamline I10 at the Diamond Light Source, Harwell, UK, and employed the RASOR diffractometer\cite{beale2010rasor_410} (preliminary measurements were made on both I10 and I06, Diamond Light Source). In order to both apply and accurately measure magnetic fields at the sample position, a custom designed sample holder (fig.~\ref{Hexaferrite_Paper_SampleMount}) was built, allowing up to two square neodymium permanent magnets to be housed underneath the sample positioned such that their magnetic field lies within the plane of the sample [fig.~\ref{Hexaferrite_Paper_Structures}(a)]. The sample holder also contains an Arepoc HHP-NU Hall sensor to allow precise recording of the applied field throughout the experiment. Rigidly attached to the sample stage is a soft iron core electromagnet, supplied by a high current source, which is aligned with the field of the permanent magnets and enables fine adjustments to be made to the field they generate.

\begin{figure}[h!]
\begin{center}
\includegraphics[scale=0.975]{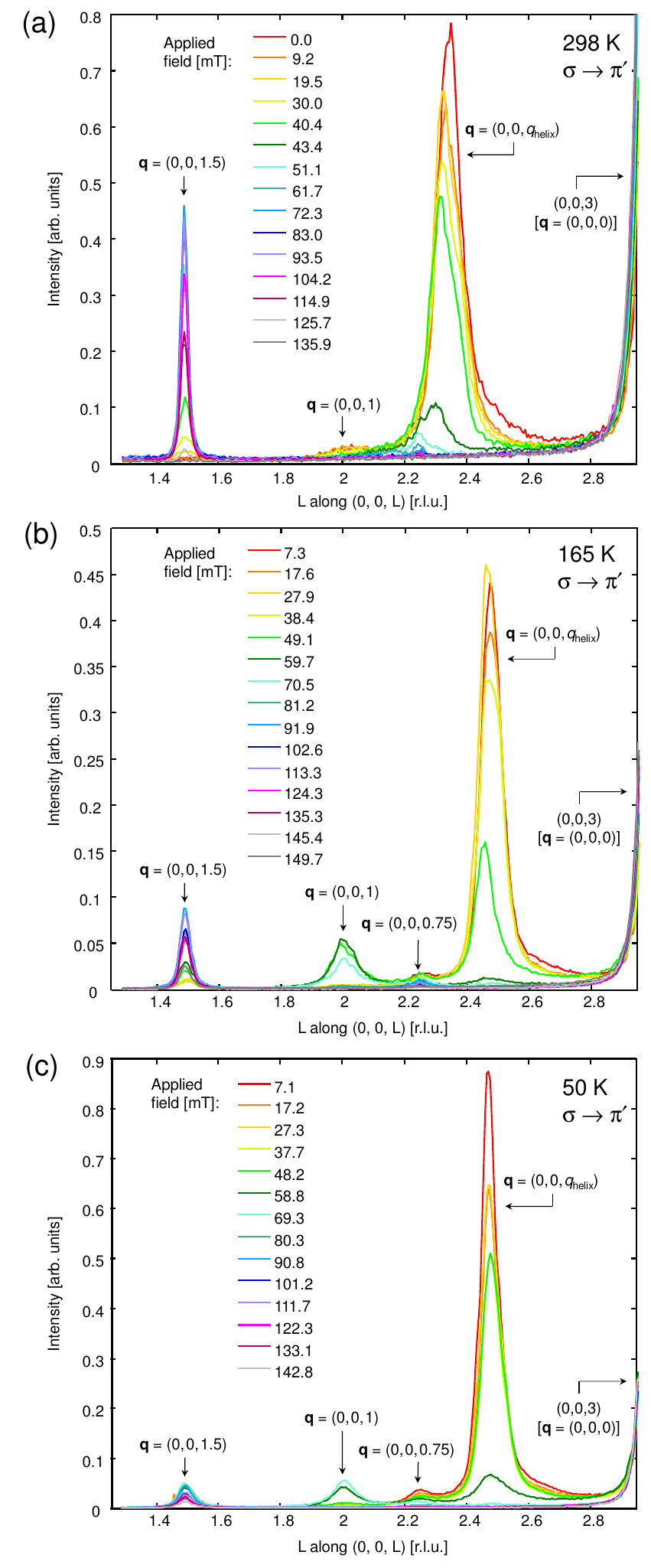}
\caption{(Color online) $l$-scans at various magnetic fields, taken in the $\sigma \rightarrow \pi^\prime$ channel at (a) 298~K, (b) 165~K, and (c) 50~K. Peaks are labelled with the corresponding propagation vectors $\mathbf{q}$, and the position of the $(0,0,3)$ Bragg peak is indicated. The integrated intensities are given in appendix \ref{Ap:IntegratedIntensities}. \label{Hexaferrite_Paper_allLscans}}
\end{center}
\end{figure}

The energy dependence of the intensity of magnetic satellites originating from the zero field incommensurate helical structure was measured and a large resonant enhancement was observed at 708.6~eV (this energy was used throughout the rest of the experiment: the X-ray attenuation length at this energy is $\approx 0.5~\mu\textrm{m}$). In order to suppress effects due to charge scattering and specular reflections arising from this scattering geometry, the measurements of the diffracted intensity were made using linearly polarised X-rays in the $\sigma \rightarrow \pi^\prime$ channel [fig.~\ref{Hexaferrite_Paper_Structures}(a)], using an analyser multilayer whose $d$-spacing (interlayer spacing) is appropriate for use at the iron edge. The resonant magnetic scattering amplitude is given by
\begin{equation}
F_{\textrm{RES}}^{\textrm{mag}}(\mathbf{Q}) = \sum_j f_{j,\textrm{RES}} \: e^{i \mathbf{Q} \cdot \mathbf{r}_j} 
\label{eq:Fres}\end{equation}
where, for $\sigma \rightarrow \pi^\prime$ scattering,
\begin{equation}
f_{j,\textrm{RES}} = -iF^{(1)}  \boldsymbol{\mu}_j \cdot (\hat{\mathbf{e}}_x \cos \theta - \hat{\mathbf{e}}_z \sin \theta) 
\label{HEXA_eqn:scatteringMatrix}\end{equation}
(the term giving rise to second order magnetic satellites \cite{hannon1988x_148} is not considered here because no such satellites were observed in the diffraction from the  incommensurate helical structure). In this expression the magnetic moment of the $j$th L or S block is written $\boldsymbol{\mu}_j$, $\theta$ is half of the scattering angle, and the basis vectors $\hat{\mathbf{e}}_x$ and $\hat{\mathbf{e}}_z$ are shown in fig.~\ref{Hexaferrite_Paper_Structures} (note that if the magnetic field is perfectly aligned along $\hat{\mathbf{e}}_x$ then due to the geometry of the polarisation analysis this technique is not sensitive to the $y$-component of the magnetic structure, which would  imply that modulations in this direction ought not to be observed: see table \ref{HEXA_tab:Satellites}). $F^{(1)}$ is a constant provided the incident wavelength is unchanged. The incident X-rays are set to be $\sigma$ polarised by the soft X-ray APPLE II undulator (the polarisation is close to 100\% \cite{wang2012complete_407}), and the crosstalk from the polarisation analyser multilayer was measured on the direct beam to be smaller than $1\%$. Since all of the magnetic structures propagate along the $\mathbf{c}$ direction, $\theta-2\theta$ scans were made along $(0,0,l)$ at various temperatures and magnetic fields. These scans contain the Bragg peak at $(0,0,3)$ (Bragg peaks are subject to $-h+k+l = 3n$ and this is therefore the only charge peak that is observed with the present setup) and magnetic satellite peaks around this. 

\begin{figure}[tb]
\begin{center}
\includegraphics[scale=0.65]{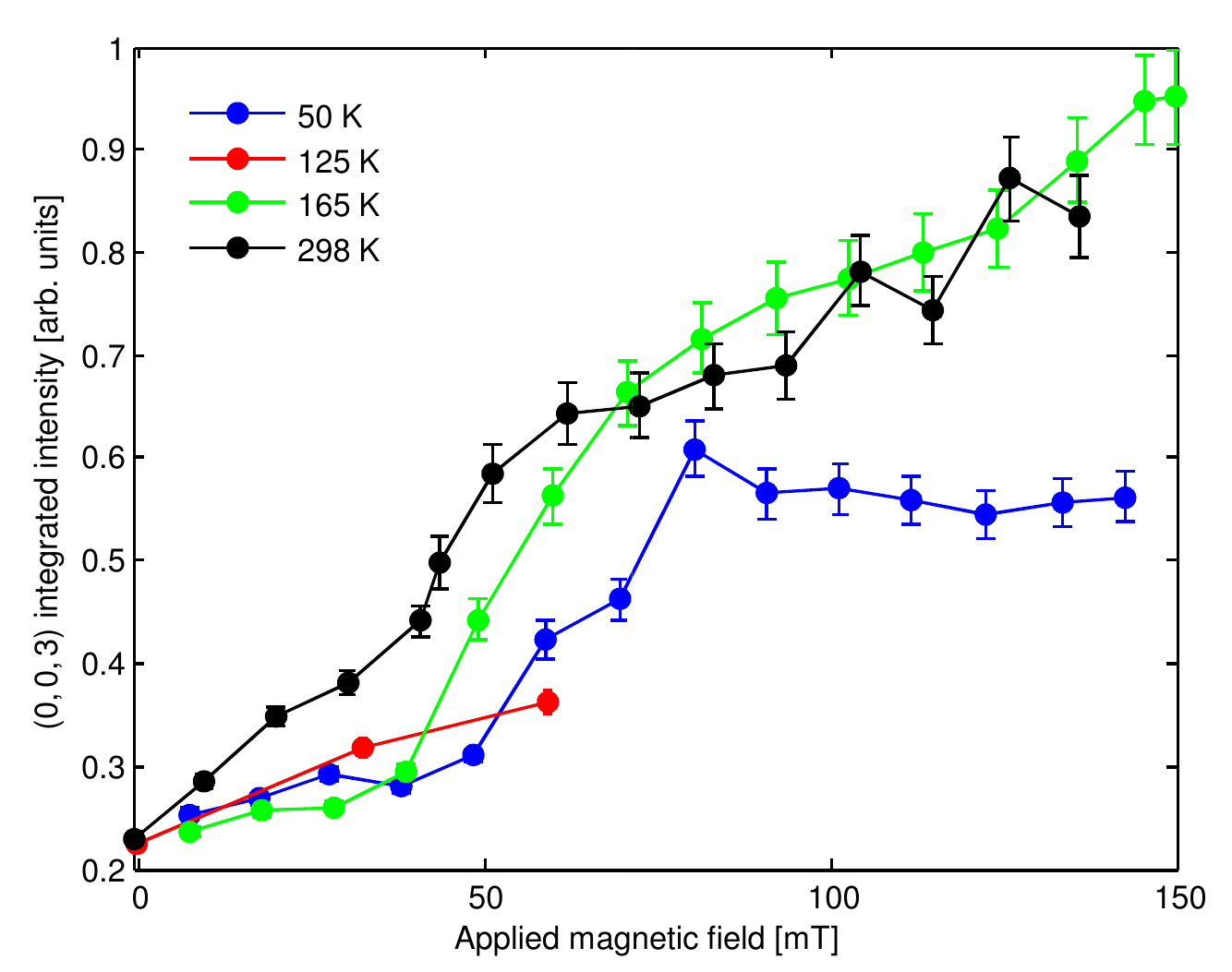}
\caption{(Color online) Intensity of the $(0, 0, 3)$ peak measured in $\sigma \rightarrow \pi^\prime$ as a function of applied magnetic field at different temperatures.\label{Hexaferrite_Paper_003intensities}}
\end{center}
\end{figure}

\section{Results}
\noindent
Fig.~\ref{Hexaferrite_Paper_allLscans} shows a detailed magnetic field dependence of the positions and intensities of the magnetic satellites from the $l$-scans [on the left of the $(0,0,3)$ Bragg peak only: satellites on the right appear similar] at 298~K, 165~K  and 50~K. In each case the sample was cooled to the desired temperature in zero magnetic field, and the applied magnetic field was increased step by step as the diffraction measurements were made. The data were all collected in the $\sigma \rightarrow \pi^\prime$ channel. The room temperature data [fig.~\ref{Hexaferrite_Paper_allLscans}(a)] show a strong contribution from the helix at $\mathbf{q} = (0,0,q_\textrm{helix}) \approx (0,0,0.65)$, which diminishes with applied field and is almost entirely gone by 50~mT. As the helical phase disappears, a peak onsets with $\mathbf{q} = (0,0,1.5)$ and is strongest at $\approx80$~mT. At higher fields this peak then also decreases leaving no magnetic satellites above 130~mT, although there is a significant increase in the intensity of the $(0,0,3)$ Bragg peak as the ferrimagnetic component develops -- see fig.~\ref{Hexaferrite_Paper_003intensities}. In addition there are very small intensities measured at peaks with propagation vectors $\mathbf{q} = (0,0,1)$ and $\mathbf{q} = (0,0,0.75)$ to be found in the room temperature data. The data taken at 165~K [see fig.~\ref{Hexaferrite_Paper_allLscans}(b)] are different from the room temperature data in three main ways: (i) they show a marked increase in the intensity of the $\mathbf{q} = (0,0,1)$ peak over several of the intermediate fields; (ii) the propagation vector of the helix has changed to be much closer to $\mathbf{q} = (0,0,0.5)$, making the very small $\mathbf{q} = (0,0,0.75)$ peak more obvious; and (iii) at the higher end of the range of fields studied here there are still satellites present with $\mathbf{q} = (0,0,1.5)$, in contrast to the room temperature data. Data taken at 50~K [fig.~\ref{Hexaferrite_Paper_allLscans}(c)] are similar to the data taken at 165~K, although the peak corresponding to the helical phase appears stronger in low fields, relative to the other magnetic satellites, than in the 165~K data set. In addition the satellites measured here with $\mathbf{q} = (0,0,1.5)$ are slightly weaker than at higher temperatures.

Having integrated the peak intensities it is possible to compare the relative sizes of each peak as a function of applied field and temperature. This comparison is made in fig.~\ref{Hexaferrite_Paper_PhaseDiagram}, which shows the sizes of the peaks with helical [\emph{i.e.~}$(0,0,q_\textrm{helix})$], $(0,0,1)$, and $(0,0,1.5)$ propagation vectors as a scatter plot (the size of each data point is scaled relative to the most intense measurement of the peak in question). Once this is done it becomes clear that the phase diagram can be split approximately into four regions according to the different propagation vectors. There are some similarities to the phase diagram of fig. 1(b) in ref.~\onlinecite{kimura2005electric_24}, for example the helical phase disappearing at approximately 50 mT with the onset of the 4-fan phase and, at higher fields, the 2-fan phase. However, our data show a greater degree of phase coexistence. The region labelled $ (0,0,0) $ corresponds to those fields and temperatures where no satellites are seen and magnetic intensity is observed purely at the ferrimagnetic position $ \mathbf{q} = (0,0,0) $. It should be emphasised that this is very much a qualitative treatment, used in order to determine the approximate behaviour of the system and identify those areas of phase coexistence. Therefore the `phase boundaries' indicated in fig.~\ref{Hexaferrite_Paper_PhaseDiagram} with the dashed lines are only suggested (approximate) positions. The weak $ \mathbf{q} = (0,0,0.75) $ peaks observed in the data imply that there is a slight deviation in the direction of the applied magnetic field from within the scattering plane, enabling one to see a small projection of the $ \mathbf{q} = (0,0,0.75) $ component from the 4-fan structure [fig.~\ref{Hexaferrite_Paper_Structures}(d)]. This deviation will be quantified in the following to allow the measured intensity profiles to be calculated. 

\section{Discussion and calculations}
\noindent
\begin{figure}[tb]
\begin{center}
\includegraphics[scale=0.4]{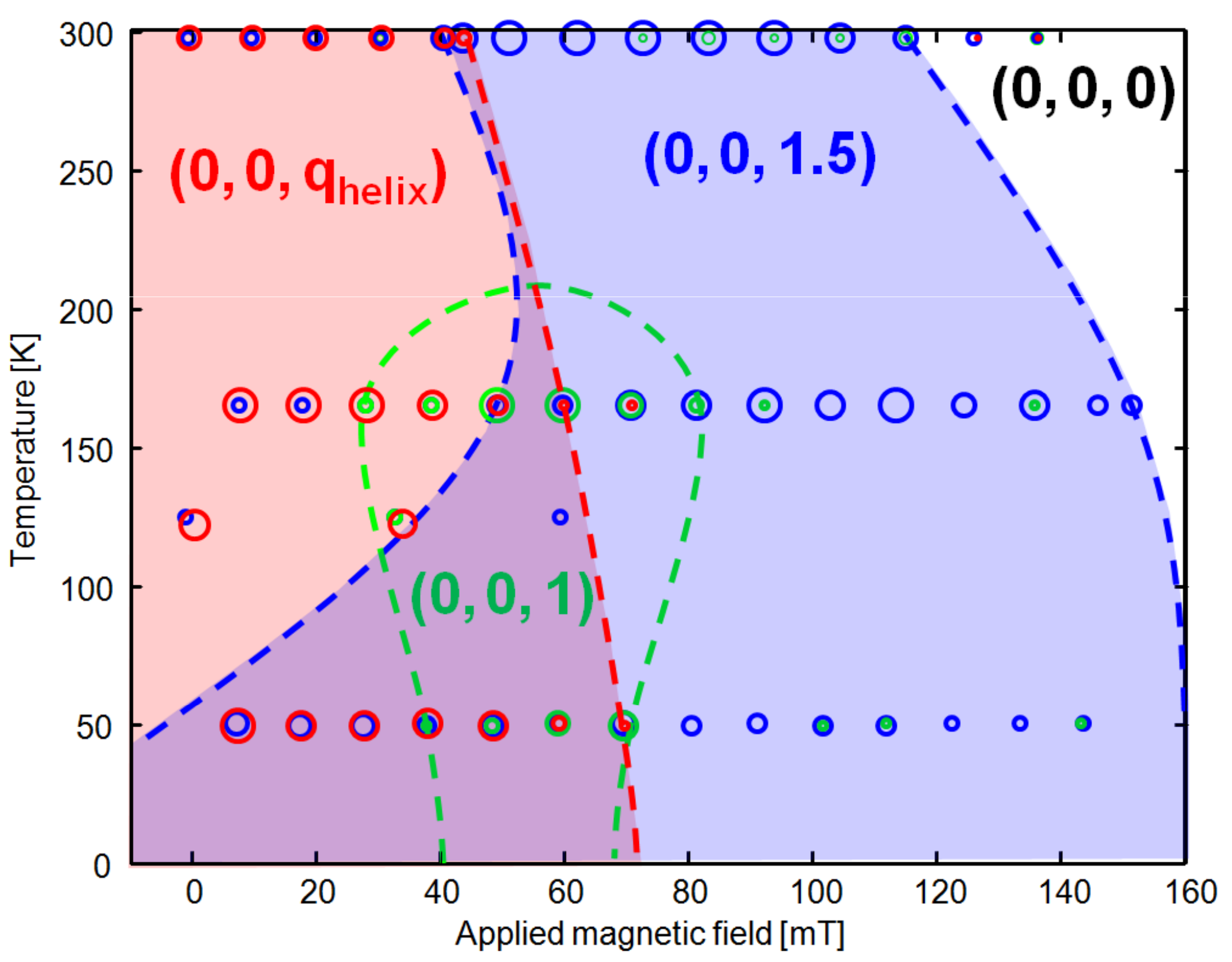}
\caption{(Color online) Low field phase diagram based on the diffraction data including that of fig.~\ref{Hexaferrite_Paper_allLscans}. The points are coloured according to the propagation vector of the diffraction peak they represent, and sized according to their relative intensities. \label{Hexaferrite_Paper_PhaseDiagram}}
\end{center}
\end{figure}

Based upon the above data, the main findings are:
\begin{enumerate}
\item
As the field is increased at room temperature, the system moves from a helical phase via the 4-fan phase with $\mathbf{q} = (0,0,1.5)$ to the 2-fan phase with $ \mathbf{q} = (0,0,0) $. There is significant phase coexistence between the helical and 4-fan phase, as evidenced by the small $(0,0,1.5)$ peaks that are present at low fields, and the growing $ \mathbf{q} = (0,0,0) $ contribution with field can be seen at the $(0, 0, 3)$ peak position (fig.~\ref{Hexaferrite_Paper_003intensities}). This contribution does not appear to saturate in the range measured here, indicating that the angle $ \phi_2 $ of the 2-fan structure is continuing to reduce at the highest measured fields and that the system has not reached a pure ferrimagnetic phase.
\item
At intermediate temperatures (165~K) the system behaves similarly, with the addition of a new peak at $\mathbf{q} = (0,0,1)$ that appears close to the region between the helical and 4-fan phases. The phase responsible for this peak has not been previously reported.
\item
At lower temperatures, evidence of the new phase with $\mathbf{q} = (0,0,1)$ is still present, but the fact that the intensity of the $(0, 0, 3)$ peak levels out above $\approx 70~\textrm{mT}$ (see fig.~\ref{Hexaferrite_Paper_003intensities}), and at a much lower value than is observed in the higher temperature data, suggests that the 2-fan phase has not been stabilised and one is instead observing the ferrimagnetic component belonging to the 4-fan structure, whose angle $ \phi_4 $ does not appear to be changing with field.
\item
Generally the measurements show a good deal of phase overlap (due to different parts of the crystal stabilising into different phases), suggesting that all of the observed phases are very close in energy.
\end{enumerate}
A magnetic `6-fan' structure that explains the peaks at $\mathbf{q} = (0,0,1)$ is shown in fig.~\ref{Hexaferrite_Paper_Structures}(c) and discussed further below. It is straightforward to calculate the exchange energies per $\textrm{S}+\textrm{L}$ block of the \textit{known} magnetic arrangements, and these are given by\cite{momozawa1993field_26}
\begin{align}
E_\textrm{helix} &= -2J_\textrm{LS}S_\textrm{S} S_\textrm{L} \cos\left(\frac{\phi_0}{2}\right) + (J_\textrm{SS}S_\textrm{S}^2 + J_\textrm{LL}S_\textrm{L}^2) \cos \phi_0 \label{HEXA_eqn:HelixEnergyPerBlock}, \\
E_\textrm{4-fan} &= J_\textrm{LL} S_\textrm{L}^2 \cos \phi_4 + \frac{J_\textrm{SS}}{2} S_\textrm{S}^2 \cos \phi_4 + \frac{J_\textrm{SS}}{2} S_\textrm{S}^2  \nonumber\\ 
& \quad{}- 2 J_\textrm{LS} S_\textrm{L} S_\textrm{S} \cos\left(\frac{\phi_4}{2} \right) - \frac{1}{2} g \mu_\textrm{B} B S_\textrm{L} (1 + \cos \phi_4) \nonumber\\
& \quad{}+ g \mu_\textrm{B} B S_\textrm{S} \cos \left(\frac{\phi_4}{2} \right), \\
E_\textrm{2-fan} &= -2 J_\textrm{LS} S_\textrm{S} S_\textrm{L} \cos \left(\frac{\phi_2}{2}\right) + J_\textrm{LL} S_\textrm{L}^2  \cos \phi_2 \nonumber\\
& \quad{}+ J_\textrm{SS} S_\textrm{S}^2 - g \mu_\textrm{B} B S_\textrm{L} \cos \left(\frac{\phi_2}{2}\right)   + g \mu_\textrm{B} B S_\textrm{S},  \\
E_\textrm{ferri} &= -2J_\textrm{LS} S_\textrm{S} S_\textrm{L} +   J_\textrm{LL} S_\textrm{L}^2 \nonumber\\
& \quad{}+ J_\textrm{SS} S_\textrm{S}^2 - g \mu_\textrm{B} B S_\textrm{L} + g \mu_\textrm{B} B S_\textrm{S},
\end{align}
where $J_{ij}$ is the exchange constant between neighbouring blocks of types $i$ and $j$, with $i,j = \textrm{L},\textrm{S}$, and the block moments have magnitudes $|\boldsymbol{\mu}_j| = g \mu_\textrm{B} S_{j}$. Higher order exchange paths will not be considered here. Similarly, the energy of the proposed 6-fan structure is given by
\begin{align}
E_\textrm{6-fan} &= J_\textrm{LL} S_\textrm{L}^2 \cos \phi_6 + \frac{2}{3} J_\textrm{SS} S_\textrm{S}^2 \cos \phi_6 + \frac{1}{3} J_\textrm{SS} S_\textrm{S}^2 \nonumber\\
& \quad{}- 2J_\textrm{LS} S_\textrm{L} S_\textrm{S} \cos \left(\frac{\phi_6}{2} \right) + \frac{g \mu_\textrm{B} B S_\textrm{S}}{3}\left[1 + 2 \cos \phi_6 \right] \nonumber\\
&\quad{}- \frac{g \mu_\textrm{B} B S_\textrm{L}}{3} \left[ \cos\left(\frac{3 \phi_6}{2} \right) + 2 \cos \left(\frac{\phi_6}{2} \right) \right] .
\end{align}

\begin{figure}[t]
\begin{center}
\includegraphics[scale=0.5]{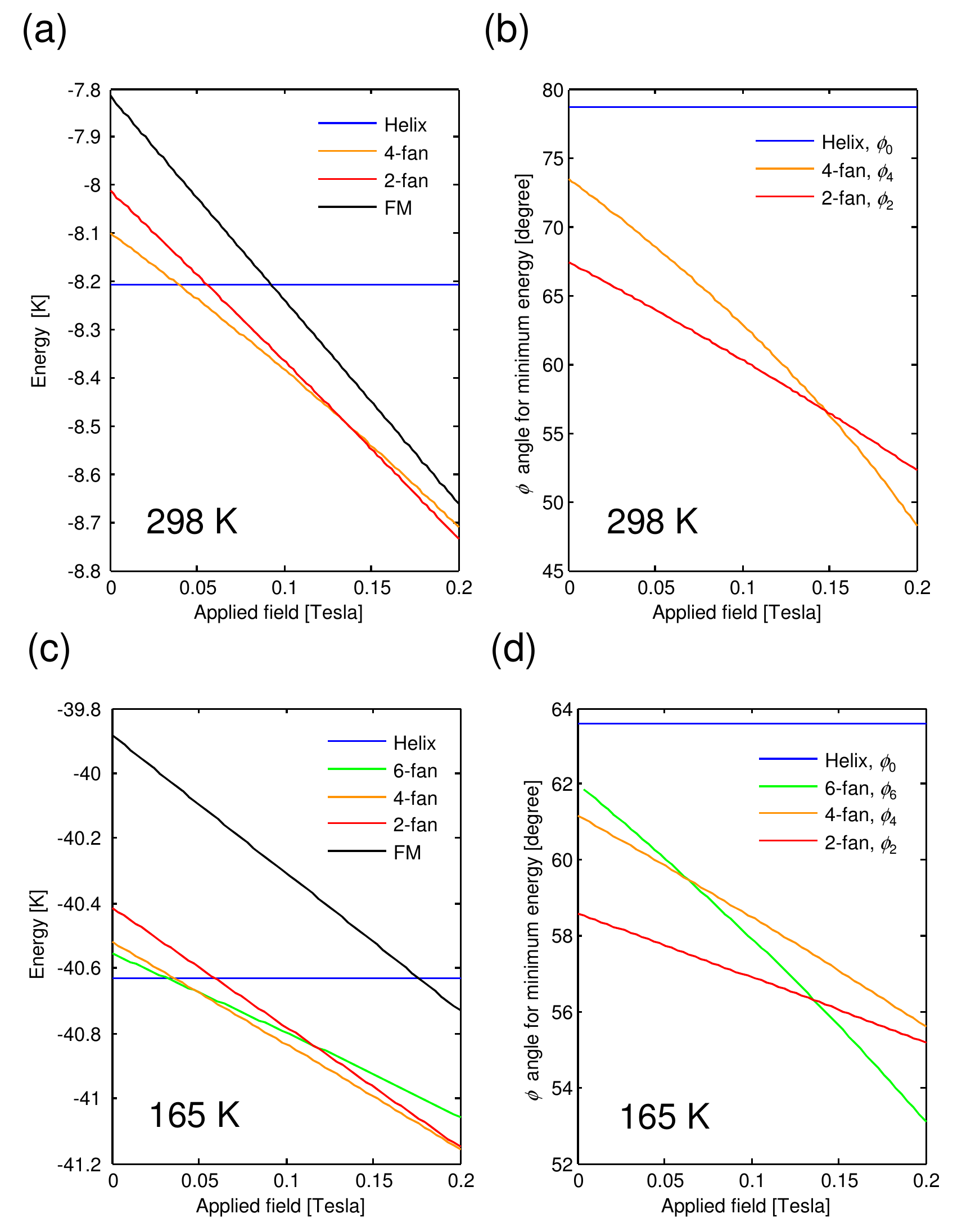}
\caption{(Color online) (a) Calculated energies (in temperature units) of the various structures as a function of field at 298~K. (b) The angles associated with each structure as a function of field. Equivalent calculations appropriate for the behaviour at 165~K (now including the 6-fan structure) are shown in (c) and (d). 
 \label{Hexaferrite_Paper_calculations}}
\end{center}
\end{figure}

Considering first the behaviour at room temperature, the three exchange constants can be determined by the size of the turn angle of the helix in zero field [$\mathbf{q}_\textrm{helix} = (0,0,0.656)$ implies $\phi_0 = 78.72^{\circ}$], together with imposing that the helical and 4-fan structures have the same energy at 40~mT (the approximate position of the phase boundary) and that the 4-fan and 2-fan phases are of the same energy at 130~mT (see fig.~\ref{Hexaferrite_Paper_PhaseDiagram}). This results in 
\begin{equation}
J_\textrm{LS} = 6.7~\textrm{K}, \:\:\:\:\: J_\textrm{SS} = 4.1~\textrm{K},  \:\:\:\:\: J_\textrm{LL} = 0.3~\textrm{K},
\end{equation}
and the block moments, determined from magnetisation measurements on the same sample, are
\begin{equation}
S_\textrm{L} = 3.4014, \quad{}\quad{} S_\textrm{S} = 0.2534. 
\end{equation}
The positive signs of the exchange constants show that every interaction is antiferromagnetic in nature. Figs.~\ref{Hexaferrite_Paper_calculations}(a) and \ref{Hexaferrite_Paper_calculations}(b) plot the calculated energy of the phases (helix, 4-fan, 2-fan, and ferrimagnetic) as well as the angles $\phi_0$, $ \phi_4 $, and $\phi_2$, as a function of field. This clearly shows that the helical phase is stabilised below 40~mT. Above this field the 4-fan structure has the lowest energy, but as the field tends towards 130~mT the energy of the 2-fan structure tends to the same value. The ferrimagnetic structure, by comparison, remains significantly higher in energy than the other phases, confirming that the ferrimagnetic phase is not stabilised until fields substantially higher than those used in the present experiment are applied. 

The 165 K data show that the helical phase turn angle has now reduced to $\phi_0 = 63.6^\circ$, from which (as above) a constraint on the exchange constants can be inferred. Since the 6-fan phase exists alongside the other phases, in a simple model another condition can be obtained by requiring that the minimum energy of the 6-fan phase occurs in the middle of the region it occupies in the phase diagram: this is at $\approx 50$~mT. The third condition used here is that the exchange constants $J_\textrm{LS}$ and $J_\textrm{LL}$ scale in the same way with temperature (this is explained in ref.~\onlinecite{momozawa1993field_26}) whereas $J_\textrm{SS}$ scales differently. Applying these three conditions gives the following three exchange constants:
\begin{equation}
J_\textrm{LS} = 32.7~\textrm{K}, \:\:\:\:\: J_\textrm{SS} = 6.6~\textrm{K},  \:\:\:\:\: J_\textrm{LL} = 1.5~\textrm{K}.
\end{equation}
Thus, relative to the room temperature values, $J_\textrm{LS}$ (and therefore $J_\textrm{LL}$) have increased by factors of $\approx 5$, whereas $J_\textrm{SS}$ has increased by a factor of only 1.6. Whilst such a large temperature dependence is unusual (and may be of interest for further studies) it is consistent with the findings of Momozawa \textit{et al.}\cite{momozawa1993field_26}, in which $J_\textrm{LL}$ and $J_\textrm{LS}$ are reported to decrease with temperature down to $\approx 77~\textrm{K}$, but $J_\textrm{SS}$ remains constant below $\approx 280~\textrm{K}$. Figs.~\ref{Hexaferrite_Paper_calculations}(c) and \ref{Hexaferrite_Paper_calculations}(d) show the energies and angles of all the different phases as a function of applied field. It is clear from this that the 6-fan structure is the minimum energy configuration between $\approx$~30~mT and 50~mT, after which the competing 4-fan structure becomes slightly lower in energy. However, in the lowest field regions there is very little difference between the energies of these phases, so it is reasonable that the diffraction data show a peak corresponding to the 6-fan structure up until $\approx 75~\textrm{mT}$. This analysis also shows that towards the upper end of the phase diagram, with fields of approximately 200~mT, the 2-fan phase becomes most stable. The fact that this model allows for a 6-fan phase to be stabilised confirms that a structure of this type is a good candidate for the hexaferrite at low fields, in order to give rise to the observed peaks with propagation vector $\mathbf{q} = (0,0,1)$ in the diffraction. In addition, the energy of the 6-fan structure at room temperature is always greater than that of one of the other structures (either helical, 4-fan, or 2-fan) which explains why the 6-fan is not observed in the room temperature data. More generally, these calculations confirm that, as in all magnetically frustrated systems, there exist a large number of states which are all very close in energy to that of the ground state (indeed, other similar systems might be expected to show longer-wavelength fan structures, such as an 8-fan or higher). This explains why the phase diagram shows large regions of phase coexistence (for example, at 50~mT and 165 K signatures of the helical, 6-fan, \textit{and} 4-fan phases are all present in the diffraction data) and why a striking variety of different orderings appears in this system.  

\begin{figure}[h!]
\begin{center}
\includegraphics[scale=0.8]{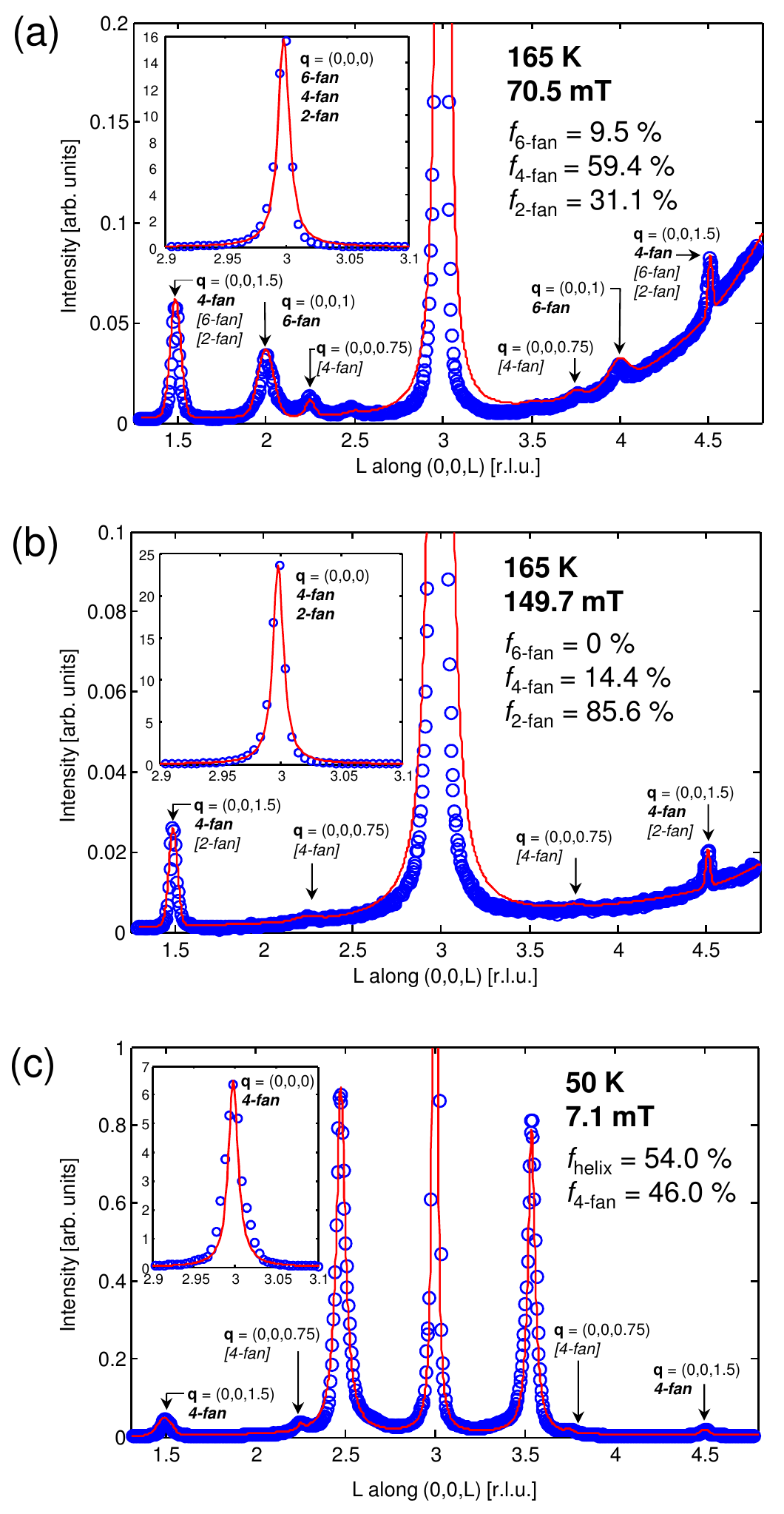}
\caption{(Color online) Measured (blue circles) and calculated (red lines) intensities for three $l$-scans: two at 165~K and one at 50~K. Data were obtained in the $\sigma \rightarrow \pi^\prime$ channel. In (a) the field is 70.5~mT and the intensities are calculated for a three-phase model (comprising 6-fan, 4-fan, and 2-fan structures). The phase fractions are indicated. All three phases contribute intensity to the $(0,0,3)$ peak, and the 4-fan and 6-fan structures result in magnetic satellites as indicated. In (b) the field is 149.7~mT and only the 4-fan and 2-fan structures remain. In (c) the low field data at 50~K are shown, and can be fitted with a model that assumes only helical (peaks at $l \approx 2.5$ and $l \approx 3.5$) and 4-fan (labelled) structures. The insets show the full Bragg peak measured in each scan with the associated calculations.
\label{Hexaferrite_Paper_Simulated_Intensities}}
\end{center}
\end{figure}

Having established the 6-fan structure as a suitable candidate to explain the $\mathbf{q} = (0,0,1)$ diffraction it is possible to calculate the scattered intensities and compare these to the values measured in the experiment (we shall do this for the 165~K data as they show the full range of phases). The intensities originating from each phase are calculated separately (details are presented in appendix \ref{Ap:Calculations}). Firstly, the 6-fan structure gives contributions at $(0,0,3)$ and $(0,0,3 \pm 1)$ positions in reciprocal space, with corresponding intensities $I^\textrm{6-fan}_{(0,0,3)}$ [eq.~(\ref{ap_eqn:6fan003})] and $I^\textrm{6-fan}_{(0,0,3\pm 1)}$ [eqs.~(\ref{ap_eqn:6fan002}) and (\ref{ap_eqn:6fan004})]. Similarly, the 4-fan structure gives intensity at $(0,0,3)$ and $(0,0,3\pm 1.5)$, denoted $I^\textrm{4-fan}_{(0,0,3)}$ [eq.~(\ref{ap_eqn:4fan003})] and $I^\textrm{4-fan}_{(0,0,3\pm1.5)}$ [eqs.~(\ref{ap_eqn:4fan001p5}) and (\ref{ap_eqn:4fan004p5})]. The only contribution from the 2-fan structure is at $(0,0,3)$ and the intensity is $I^\textrm{2-fan}_{(0,0,3)}$ [eq.~(\ref{ap_eqn:2fan003})]. The integrated intensities of each peak can therefore be calculated via the following equations:
\begin{align}
I^{\textrm{calc.}}_{(0,0,3 \pm 1)} &= f_\textrm{6-fan} I^\textrm{6-fan}_{(0,0,3\pm 1)},\label{HEXA_eq:calcIntensity1} \\
I^{\textrm{calc.}}_{(0,0,3 \pm 1.5)} &= f_\textrm{4-fan} I^\textrm{4-fan}_{(0,0,3\pm 1.5)}, \\
I^{\textrm{calc.}}_{(0,0,3)} &= f_\textrm{6-fan}  I^\textrm{6-fan}_{(0,0,3)} +  f_\textrm{4-fan} I^\textrm{4-fan}_{(0,0,3)} + f_\textrm{2-fan} I^\textrm{2-fan}_{(0,0,3)}, \label{HEXA_eq:calcIntensity2}
\end{align}
where the $f$'s are the phase fractions (summing to unity), and a global scaling factor is to be applied between calculated and measured intensities.

Before comparing these intensities to the data, the background in the measurements (which appears to increase at high angle) is fitted empirically, and the integrated intensities measured in the data (on the low angle side of the Bragg peak) are then compared to those given by eqs.~(\ref{HEXA_eq:calcIntensity1}) to (\ref{HEXA_eq:calcIntensity2}) and the phase fractions extracted. The shapes of the magnetic satellites at both high and low angle are then fitted to the data, subject to their integrated intensities being consistent with the calculated phase fractions. The results (calculations of the intensity profiles at 165~K for two separate fields) are shown along with the measured data and the phase fractions in figs.~\ref{Hexaferrite_Paper_Simulated_Intensities}(a) and \ref{Hexaferrite_Paper_Simulated_Intensities}(b), and reproduce the intensities measured at high angle very well. Note that the data in fig.~\ref{Hexaferrite_Paper_Simulated_Intensities} show small peaks with propagation vectors $\mathbf{q} = (0,0,0.75)$: as explained above, these appear because of the $\mathbf{q} = (0,0,0.75)$ modulation in the magnetic structure of the 4-fan phase which is not precisely orthogonal to the scattering plane [the deviation in the net magnetisation direction of the 4-fan structure with respect to the scattering plane is calculated to be $\approx 5.5^\circ$ based on the relative size of the $\mathbf{q} = (0,0,0.75)$ peak in fig.~\ref{Hexaferrite_Paper_Simulated_Intensities}(a): the calculations in fig.~\ref{Hexaferrite_Paper_Simulated_Intensities} take account of this with the resulting extra contributions to the intensity being labelled in square brackets]. There is also a very weak peak in this plot at $l \approx 2.5$: this indicates that a very small part of the sample remains in the helical phase even at 70.5~mT (this was not considered in the phase fraction calculations). The calculated phase fractions give a more detailed picture of how the system as a whole responds to changes in the applied field. At 165~K and $\approx 70$~mT [fig.~\ref{Hexaferrite_Paper_Simulated_Intensities}(a)] there is a mixture of all three phases, although a relatively small part of the sample (9.5\%) appears to have stabilised in the 6-fan phase (this is because, at 70~mT, the 4-fan phase is slightly lower in energy). The majority of the sample (59.4\%) stabilises in the 4-fan phase as expected, although in order to explain the intensity measured at $(0,0,3)$ a significant portion (31.1\%) of the sample is made up of the higher energy 2-fan structure. At $\approx 150$~mT [fig.~\ref{Hexaferrite_Paper_Simulated_Intensities}(b)] none of the 6-fan phase remains, and the majority of the sample (85.6\%) exists in the 2-fan structure with the remainder stabilising into the 4-fan. This is due to the ever decreasing difference in energy between 4-fan and 2-fan structures as the applied field is increased. Fig.~\ref{Hexaferrite_Paper_Simulated_Intensities}(c) shows a fit made to the data taken at 50~K: the strong incommensurate peaks arising from the helical structure are fitted by calculating the intensities from a commensurate approximation with $\mathbf{q}_\textrm{helix} = (0,0,0.5)$ and allowing a mixture of helical (54\%) and 4-fan (46\%) phases.  %Although the fit is reasonable, the measured intensities of the $\mathbf{q} = (0,0,0.75)$ satellites (particularly at $l = 2.25$) are slightly stronger than the calculated values: we speculate that this may be due to a small part of the sample stabilising in a longer wavelength `8-fan' phase, in which the $\mathbf{q} = (0,0,0.75)$ modulation would be directed along $\hat{\mathbf{e}}_x$ and therefore be stronger than the equivalent modulation along $\hat{\mathbf{e}}_y$ present in the 4-fan structure.      

\section{Conclusion}
\noindent
Having undertaken a detailed resonant soft X-ray diffraction study of Ba$_{0.5}$Sr$_{1.5}$Zn$_2$Fe$_{12}$O$_{22}$ at low fields, the system appears to explore a multitude of phases which are all very close in energy. The helical phase is shown to exist only for the lowest fields, becoming unstable as the field is increased and other magnetic fan structures become more favourable energetically. At room temperature the system rapidly enters the 2-fan phase, whereas at lower temperatures the 4-fan phase is stable up to higher applied fields. Simple exchange energy arguments have been used to calculate the differences in energy between competing phases and, in particular, we show that a new `6-fan' phase, existing somewhere between the helical and `4-fan' (intermediate-I) structures, is a likely candidate for the previously unreported peaks in the diffraction data. Based on the calculated inter-spin-block angles in each of the commensurate magnetic structures, scattering intensities are calculated and shown to fit the data well. Phase fractions are extracted to quantify the degree of phase coexistence at 165~K and 50~K. We note that the use of a diffraction technique was crucial for the discovery of the new 6-fan phase since its net magnetisation and energy are very similar to the competing phases: thus the only real signature of its existence is a change in the periodicity of the magnetic arrangement.

% the following three lines are needed to prevent strange spacing around the acknowledegments
%\vspace*{0.5cm}
\begin{acknowledgments}
%\vspace*{-0.8cm}

\noindent  We are grateful to Radu Coldea for useful discussions and to  Mark Sussmuth and Raymond Fan for technical support during the synchrotron experiment. The work at the University of Oxford was funded by an EPSRC Grant No. EP/J003557/1, entitled `New Concepts in Multiferroics and Magnetoelectrics'. The work at Postech was supported by the Max Planck POSTECH/KOREA Research Initiative Program [No. 2011-0031558] through the NRF of Korea funded by the Ministry of Education, Science and Technology. SWC was also supported by the NSF DMR-1104484. 
\end{acknowledgments}

\bibliographystyle{apsrev}
\bibliography{HEARMON_etal_references}

\appendix

\section{Integrated intensities}\label{Ap:IntegratedIntensities}
The following tables give the integrated intensities of the satellite peaks shown in fig.~\ref{Hexaferrite_Paper_allLscans}. The intensities are in arbitrary units (the same scale is used for all three temperatures).

\begin{table}[h!]
\begin{center}
\caption{Integrated intensities of the $(0,0,1.5)$,  $(0,0,2)$, and $(0,0,3-q_\textrm{helix})$ peaks at 298~K at various magnetic fields.  \label{HEXA_tab:Appendix_298}}
\vspace{0.15cm}
\begin{tabular}{cccc}
\hline \hline \vspace{0.15cm}
Field / [mT]	&  $(0,0,1.5)$				&  $(0,0,2)$	& $(0,0,3-q_\textrm{helix})$    \\
\hline
        0   & 0.0554    & 0.0456   & 6.7573    \\
    9.2  &  0.0631  &  0.1247  &  5.9071 \\
   19.5  &  0.1092  &  0.0561   & 6.2076 \\
   30.0  &  0.2497  &  0.0194   & 5.1086 \\
   40.4  &  0.5563  &       0   & 4.3281 \\
   43.4  &  0.9891  &       0   & 0.6972 \\
   51.1  &  1.1919  &       0    &     0 \\
   61.7  &  1.4386 &       0    &     0 \\
   72.3  &  1.3938  &  0.0107    &     0 \\
   83.0  &  1.2826  &  0.0351    &     0 \\
   93.5  &  1.2839  &  0.0020    &     0 \\
  104.2  &  1.0002  &  0.0103    &     0 \\
  114.9  &  0.7526  &  0.0596    &     0 \\
  125.7  &  0.0716  &       0  &  0.0029 \\
  135.9  &  0.0082  &  0.0408  &  0.0271 \\ 
  \hline \hline
\end{tabular}\end{center}
\end{table}
\begin{table}[h!]
\begin{center}
\caption{Integrated intensities of the $(0,0,1.5)$,  $(0,0,2)$, and $(0,0,3-q_\textrm{helix})$ peaks at 165~K at various magnetic fields.  \label{HEXA_tab:Appendix_165}}
\vspace{0.15cm}
\begin{tabular}{cccc}
\hline \hline \vspace{0.15cm}
Field / [mT]	&  $(0,0,1.5)$				&  $(0,0,2)$	& $(0,0,3-q_\textrm{helix})$    \\
\hline
 7.3   & 0.6298    &     0  &   35.1153 \\
   17.6 &   0.5682 &        0  &   33.5880 \\
   27.9 &   0.5252 &   0.2573   &  38.8626 \\
   38.4 &   0.7130 &   0.1752    & 31.1416 \\
   49.1 &   1.0576 &   3.9117   &  11.4801 \\
   59.7 &   1.5435 &   4.3910   &   0.7339 \\
   70.5 &   2.9073 &   2.4653    &  0.2995 \\
   81.2 &   2.8872 &   0.1715     &      0 \\
   91.9 &   3.9298 &   0.0150    &    0 \\
  102.6 &   2.9134 &        0     &    0 \\
  113.3 &   3.6288 &        0     &    0 \\
  124.3 &   2.3503 &        0     &    0 \\
  135.3 &   2.5378 &   0.0355     &    0 \\
  145.4 &   1.1326 &        0     &    0 \\
  149.7 &   1.1603 &        0     &    0 \\
  \hline \hline
\end{tabular}\end{center}
\end{table}
\begin{table}[h!]
\begin{center}
\caption{Integrated intensities of the $(0,0,1.5)$,  $(0,0,2)$, and $(0,0,3-q_\textrm{helix})$ peaks at 50~K at various magnetic fields.  \label{HEXA_tab:Appendix_50}}
\vspace{0.15cm}
\begin{tabular}{cccc}
\hline \hline \vspace{0.15cm}
Field / [mT]	&  $(0,0,1.5)$				&  $(0,0,2)$	& $(0,0,3-q_\textrm{helix})$    \\
\hline
    7.1  &  2.7828 &         0 &   61.0975 \\
   17.2  &  2.6980 &        0  & 46.8263 \\
   27.3  &  2.7266 &        0  & 50.5775 \\
   37.7  &  2.6290 &   0.0502  & 39.5523 \\
   48.2  &  2.7552 &   0.4447  & 41.3708 \\
   58.8  &  2.3947 &   3.0498  &  5.4783 \\
   69.3  &  2.8696 &   4.3099  &  0.2730 \\
   80.3  &  2.3229 &        0  &       0 \\
   90.8  &  2.3730 &        0  &       0 \\
  101.2  &  1.6276 &   0.0231  &       0 \\
  111.7  &  1.7094 &   0.0275  &       0 \\
  122.3  &  1.0422 &        0  &       0 \\
  133.1  &  1.2196 &        0  &       0 \\
  142.8  &  0.7325 &   0.0414  &       0 \\

  \hline \hline
\end{tabular}\end{center}
\end{table}

\section{Intensity calculations}\label{Ap:Calculations}
\noindent
In this appendix we give the expressions for the scattered X-ray intensities originating from the 6-fan, 4-fan, and 2-fan phases of the hexaferrite, as used in calculating the intensity profiles in eqs.~(\ref{HEXA_eq:calcIntensity1}) to (\ref{HEXA_eq:calcIntensity2}). 

Firstly, it is necessary to determine the Fourier decomposition for the 6-fan structure (those for the 4-fan and 2-fan structures are given in ref.~\onlinecite{momozawa1993field_26}). Splitting each moment $ \mu_\textrm{L,S} $ into components along $ \hat{\mathbf{e}}_x $ and $ \hat{\mathbf{e}}_y $ (fig.~\ref{Hexaferrite_Paper_Structures}), one has for the L blocks:
\begin{align}
\mu^{x}_{\textrm{L}}(\mathbf{r}) &= A^{x}_{\textrm{L}} +  B^{x}_{\textrm{L}} \cos \left( \mathbf{c}^* \cdot \mathbf{r} + \phi^x_{\textrm{L}} \right), \\
\mu^{y}_{\textrm{L}}(\mathbf{r}) &= A^{y}_{\textrm{L}} \cos\left(\frac{1}{2}\mathbf{c}^* \cdot \mathbf{r} + \phi_\textrm{L}^y \right) \nonumber\\
&\quad{}+  B^{y}_{\textrm{L}} \cos \left( \frac{3}{2} \mathbf{c}^* \cdot \mathbf{r} + \varphi^y_{\textrm{L}} \right),
\end{align}
where $\mathbf{r} = \mathbf{0}, \frac{\mathbf{c}}{3}, \frac{2\mathbf{c}}{3}, \dots$ is the position of the (L + S) block belonging to the moment, the constants $A^{x,y}_{\textrm{L}}$ and $B^{x,y}_{\textrm{L}}$  depend on the angles of the magnetic structure in the following way:
\begin{align}
A^{x}_{\textrm{L}} &= \frac{S_\textrm{L}}{3} \cos ( {\phi_6}/{2} ) \left( 1 + 2\cos \phi_6 \right), \\
B^{x}_{\textrm{L}} &= \frac{4S_\textrm{L}}{3} \cos ( {\phi_6}/{2} ) \left( 1 - \cos \phi_6 \right), \\
A^{y}_{\textrm{L}} &= \frac{4S_\textrm{L}}{3} \sin ( {\phi_6}/{2} ) \left( \cos \phi_6 + 1 \right), \\
B^{y}_{\textrm{L}} &= \frac{2S_\textrm{L}}{3} \sin ( {\phi_6}/{2} ) \left( \cos \phi_6 - 1/2 \right),
\end{align}
and the phase angles are
\begin{align}
\phi^x_\textrm{L} = \frac{\pi}{3}, \quad{}\quad{} \phi^y_\textrm{L} = -\frac{\pi}{3}, \quad{}\quad{} \varphi^y_\textrm{L} = \pi.
\end{align} 
Similarly, the S blocks may be described by
\begin{align}
\mu^{x}_{\textrm{S}}(\mathbf{r}) &= A^{x}_{\textrm{S}} +  B^{x}_{\textrm{S}} \cos \left( \mathbf{c}^* \cdot \mathbf{r} + \phi^x_{\textrm{S}} \right), \\
\mu^{y}_{\textrm{S}}(\mathbf{r}) &= A^{y}_{\textrm{S}} \cos\left(\frac{1}{2}\mathbf{c}^* \cdot \mathbf{r} + \phi_\textrm{S}^y \right),
\end{align}
with 
\begin{align}
A^{x}_{\textrm{S}} &= -\frac{S_\textrm{S}}{3} \left( 1 + 2\cos \phi_6 \right),	\\
B^{x}_{\textrm{S}} &= \frac{2S_\textrm{S}}{3} \left( \cos \phi_6 - 1 \right),	 \\
A^{y}_{\textrm{S}} &= \frac{2S_\textrm{S}}{\sqrt{3}} \sin \phi_6,
\end{align}
and
\begin{align}
\phi^x_\textrm{S} = 0, \quad{}\quad{} \phi^y_\textrm{S} &= \frac{\pi}{2}.
\end{align}

The scattered X-ray intensities may be calculated from the square of the resonant magnetic structure factor [see eqs.~(\ref{eq:Fres}) and (\ref{HEXA_eqn:scatteringMatrix})]. Working with a supercell that contains six $(\textrm{L} + \textrm{S})$ blocks (\textit{i.e.~}containing two periods of the $\mathbf{q} = (0,0,1)$ modulation, and three periods of the $\mathbf{q} = (0,0,1.5)$ modulation), the intensities at locations $(0,0,l)$ \textit{for the 6-fan structure} are given by
\begin{align}
I^\textrm{6-fan}_{(0,0,3m)} &\propto \left| -i \cos \theta \:   \left( A_\textrm{L}^x + A^x_\textrm{S} e^{-i\pi m}  \right)  \right|^2,
\label{ap_eqn:6fan003} \\
I^\textrm{6-fan}_{(0,0,3m-1)} &\propto \left| -i \cos \theta \:    \left[ \frac{B^x_\textrm{L}}{2} e^{i\phi_\textrm{L}^x} + \frac{B^x_\textrm{S}}{2} e^{-\frac{i\pi (3m-1)}{3}}  \right]  \right|^2,
\label{ap_eqn:6fan002} \\
I^\textrm{6-fan}_{(0,0,3m+1)} &\propto \left| -i \cos \theta \:    \left[ \frac{B^x_\textrm{L}}{2} e^{-i\phi_\textrm{L}^x} + \frac{B^x_\textrm{S}}{2} e^{-\frac{i\pi (3m+1)}{3}}  \right]  \right|^2;
\label{ap_eqn:6fan004}\end{align}
those \textit{for the 4-fan structure} are given by
\begin{align}
I^\textrm{4-fan}_{(0,0,3m)} &\propto \left| -i \cos \theta \:   \left[\frac{S_\textrm{L}}{2}(1+\cos \phi_4) \right. \right. \nonumber \\
&\quad{}\quad{}\quad{}\quad{}- \left.\left. S_\textrm{S} \cos\left(\frac{\phi_4}{2} \right) e^{-i \pi m} \right]  \right|^2,
\label{ap_eqn:4fan003} \\
I^\textrm{4-fan}_{(0,0,3m-1.5)} &\propto \left| -i \cos \theta \:   \frac{S_\textrm{L}}{4}(1 - \cos\phi_4)  \right|^2,
\label{ap_eqn:4fan001p5} \\
I^\textrm{4-fan}_{(0,0,3m+1.5)} &\propto \left| -i \cos \theta \:    \frac{S_\textrm{L}}{4}(1 - \cos\phi_4)  \right|^2;
\label{ap_eqn:4fan004p5}\end{align}
and \textit{for the 2-fan structure} by
\begin{align}
I^\textrm{2-fan}_{(0,0,3m)} \propto \left| -i \cos \theta \:  \left[S_\textrm{L} \cos \left(\frac{\phi_2}{2} \right) - S_\textrm{S} e^{-i\pi m}\right]  \right|^2,
\label{ap_eqn:2fan003}\end{align}
where $m \in \mathbb{Z}$.

\end{document}